\documentclass[a4paper,14pt]{article}
\usepackage{amsmath}
\usepackage{multicol}
\usepackage{graphicx}
\usepackage{pstricks}
\usepackage{pst-plot}
\usepackage{epsfig}
\usepackage[french]{babel}
\usepackage[utf8]{inputenc}
\usepackage{graphicx}
\usepackage{color}
\usepackage{ulem}
\usepackage{multirow}
\usepackage{wrapfig}

\def\bq{\begin{quotation}}
\def\eq{\end{quotation}}

\newcommand{\ket}[1]{|\kern.3ex #1 \kern.3ex\rangle}
\newcommand{\bra}[1]{\langle\kern.3ex #1 \kern.3ex|}
\newcommand{\braket}[2]{\langle\kern.3ex #1 \kern.3ex|
                        \kern.3ex #2 \kern.3ex\rangle}
\begin{document}
\title {Les inégalités de Bell et l'article Einstein, Podolsky, Rosen,
une relecture}

\author{ P.Roussel,
Institut de Physique nucl\'eaire\\
Universit\'e Paris XI, CNRS, IN2P3 \\
F-91406 Orsay Cedex }

\date{\today}

\maketitle
\large
\section{Introduction}
 L'article de Bell\cite{bell1964} où il a proposé ses inégalités devenues célèbres,
on en verra le contenu plus loin, est peut-être encore aujourd'hui
l'article de physique le plus cité. L'article d'Einstein, Podolsky,
Rosen\cite{EPR} auquel ces inégalités sont dites être confrontées n'est pas loin
derrière, mais c'est l'article de Bell écrit 30 ans aprés qui lui a
redonné cette notoriété. Cependant, si Bell à redonné vie à l'article
EPR, c'est pour aussitôt en restreindre la portée et même en détruire
le sens.

\begin{wrapfigure}[16]{r}[34pt]{12 cm}
\centering
\includegraphics[ scale = 0.7]{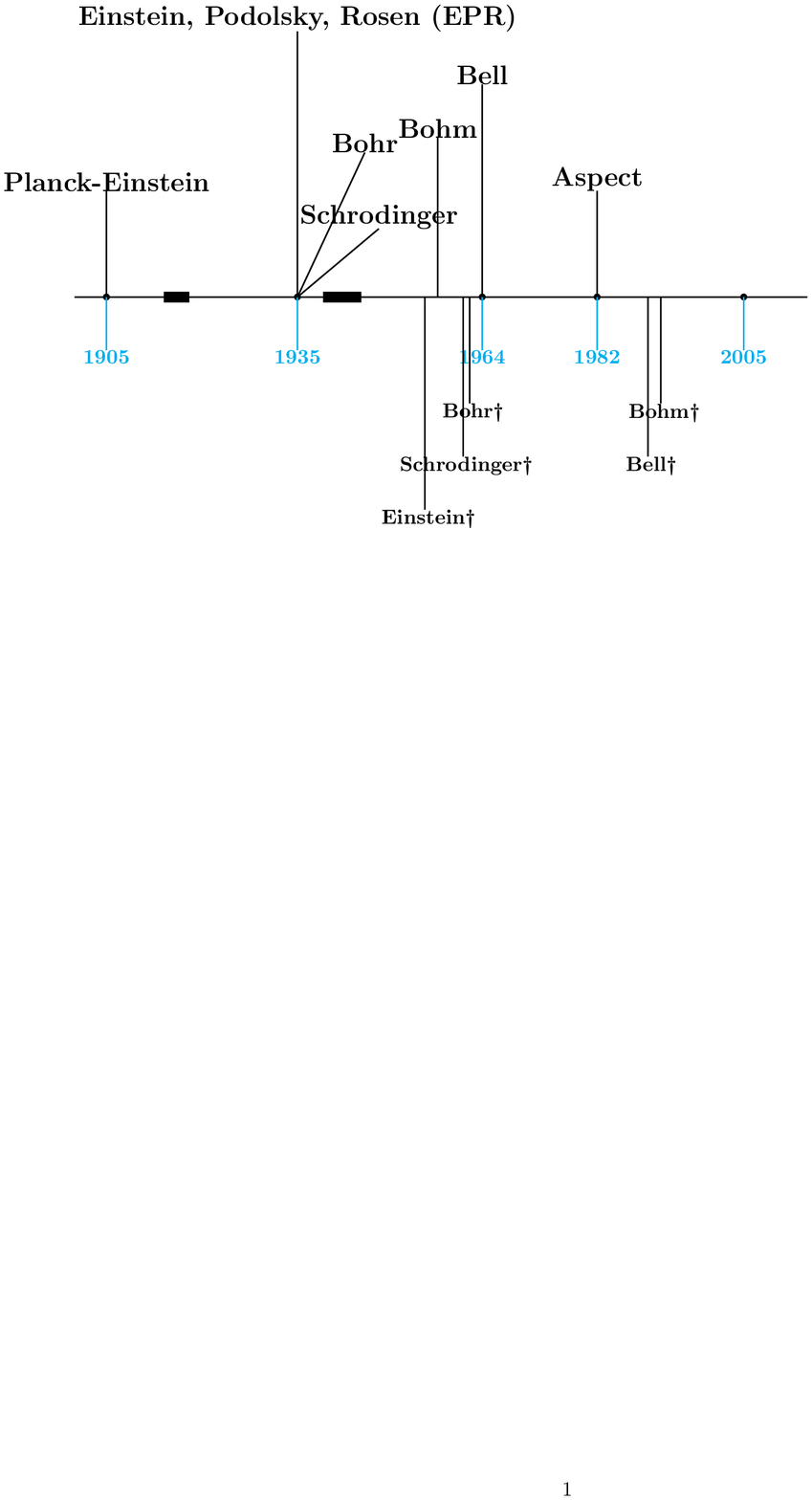}
\end{wrapfigure}

On rapporte ainsi que Bohr et Einstein étaient en désaccord sur
l'interprétation de la mécanique quantique mais que Bell avec ses
inégalités a permis de trancher, et que l'expérience a donné raison
{\sl définitivement} à Bohr après qu' aient été confirmée la violation
attendue de ces inégalités, à partir de 1971 et plus fortement en 1982
avec les expériences d'Alain Aspect\cite{Aspe82}.

Notons que si l'article de Bell atteint un record de popularité, on
pourrait dire aussi qu'il atteint un record de portée dans le temps :
écrit en 1964 à propos d'un article de 1935, il conduit à des
expériences dans les années 1970-1980 et il est encore souvent cité
aujourd'hui en 2014. Une histoire singulière dans le monde
scientifique.
 
En reprenant de plus près les textes originaux, on va tenter de
retrouver l'enchainement des arguments et examiner s'ils se laissent
bien contenir dans cette histoire dont le calendrier étendu sur un
demi-siècle est présenté sur la figure 1 ci-dessus où sont indiquées
en haut les interventions importantes sur le sujet et en bas, la
disparition des intervenants.

\section{EPR et Bell, un début tout en prudence}

Commençons par le titre de  l'article de Bell\footnote{le texte de
Bell est imprimé en bleu pour faciliter la lecture} :

\bq
{\color{blue}\sl On the Einstein Podolsky Rosen paradox}.
\eq

On remarquera la prudence de ce titre qui n'annonce pas du tout la
confrontation qui va suivre. Juste évoque-t-il le sujet, l'article de
Einstein, Podolsky et Rosen, et le qualifie-t-il de paradoxe. Pourquoi
cette prudence et pourquoi ce qualificatif\footnote{ Pour le TLF :
`Affirmation surprenante en son fond et/ou sa forme, qui contredit les
idées reçues, l'opinion courante, les préjugés''}? Et confrontons le
titre de Bell avec celui de Einstein, Podolsky, Rosen\footnote{les
textes issus de EPR ou d'Einstein sont imprimés en bleu et
souslignés}:
 
\bq
{\color{blue}\sl\uline{ Can quantum-mechanical description of physical reality be
considered complete?}}
\eq

Une grande prudence donc là aussi, une simple question, et peut-être
cette prudence du titre choisie par EPR, il faudra comprendre
pourquoi, explique-t-elle celle de Bell... dans son titre!
Quant au paradoxe, une question ne peut constituer un paradoxe,
éventuellement la réponse.

Et en effet, la réponse est là chez EPR, dès la fin du résumé :

\bq
{ \color{blue}\sl\uline{ One is thus led to conclude that the description of reality as
given by a wave function is not complete}}.
\eq

Il y a bien là de quoi ``contredire l'opinion courante'', ``les
préjugés'', on peut bien parler de paradoxe, même si le mot n'est pas
chez EPR, à cette nuance près qu'il ne s'agit pas d'une affirmation
mais bien d'une démonstration.Il restera à voir précisément de quoi.

Même si la réponse du résumé est plus précise que la question du titre
(c'est le caractère exhaustif de la fonction d'onde qui est remis en
cause), on doit s'interroger sur la raison qui a conduit Einstein et
ses collègues à choisir le mode interrogateur pour le titre de
l'article alors que le texte comme le résumé défendent une réponse si
clairement négative. Pourquoi également le faire porter sur la notion
la plus générale de { \sl description de la réalité physique}. On y
reviendra.

\section{EPR de plus près.}

Avant de poursuivre avec Bell, il devient nécessaire d'analyser plus
en détail l'article EPR. De distinguer ce qui est démonstration de ce
qui est commentaires. Une démonstration est éternelle les commentaires
dépendent de l'époque, du contexte, des questions qui s'y posaient.

L'article pose d'abord les 2 conditions qui fondent pour eux (pour
nous) le succès d'une théorie physique.{\bf 1) La théorie est-elle
correcte, ses prédictions conformes à la réalité des observations. 2)
Est-elle complète c.a.d. tout élément de réalité trouve-t-il un
correspondant dans la théorie.}

EPR écrivent explicitement que c'est la seconde question qui est
traitée dans leur article. Bien que ce ne soit pas dit, on peut penser
que l'examen de la question 2) suppose acquise la réponse positive à
la question 1). C'était en tout cas l'avis d'Einstein exprimé de
multiples fois bien avant 1935 comme bien après(voir plus loin). 
L'article propose ensuite un long exposé pour associer, en général,
pour un système physique isolé, la réalité d'une grandeur pour ce
système avec la prédictabilité de cette grandeur. Pourquoi cette
insistance, c'est que la mécanique quantique à laquelle EPR se
tiennent rigoureusement ne prétend pas trouver la réalité des choses
mais seulement prédire le résultat d'une mesure ou la distribution des
probabilités de ses résultats.  A une grandeur dont le résultat de la
mesure est prédictible, EPR font donc correspondre une réalité
physique. La mécanique quantique ne s'oppose pas à cela.

Insistons que EPR, eux, et au contraire de leurs contemporains parlent
bien de ``réalité objective'' fortement contestée à l'époque (encore
aujourd'hui!) mais ils reprennent et utilisent fidèlement les
préceptes de la mécanique quantique. Ceux de leur époque, les mêmes
qu'aujourd'hui.

\begin{itemize}

\item 
Le concept d'état caractèrisé par une fonction d'onde qui fournit
toute l'information disponible sur l'état.
\item
Le concept d'observable et son opérateur correspondant.
\item
La particularité d'un état propre qui seul permet de prédire avec
certitude la valeur de l'observable correspondante.
\item
Le processus de ``mesure'' (c'est clairement pour EPR un processus,
aucune tentation pour l'appel à la conscience de l'observateur!), le
changement d'état (réduction du paquer d'onde) auquel il conduit en
général (voir appendice A) par une interraction incontrôlable avec l'appareil de mesure.
\item
Enfin, l'impossibilité pour un état d'être à la fois état propre de
deux opérateurs qui ne commutent pas, position et impulsion par
exemple, mais c'est celui là qui sera utilisé plus loin : Un état
propre de $X$ permet de connaitre avec certitude x mais ne dit rien,
sinon une distribution de probabilité, sur la variable conjuguée$
P_x$.
\end{itemize}

Tout cet arsenal est mis en route par EPR, mais il est utilisé dans le
cas particulier où le système étudié est composé de deux éléments. Là
est l'originalité du cas examiné et la surprise du résultat
démontré. Disons tout de suite que la réponse démontrée est une
alternative et dont les deux termes sont embarassants, là est la
subtilité. Embarassants pour EPR et embarassants pour (presque!) tout
le monde. On verra tout de même que l'un plus que l'autre. Et c'est de
cet ambarras général que peut résulter finalement la conclusion que la
mécanique quantique n'est pas complète. Le titre d'EPR pose au moins
la question. La démonstration va aller plus loin.

\section{Le coeur de la démonstration : l'alternative EPR}

La démonstration est à trouver dans l'article original bien sûr, mais
on peut en donner ici la trame du déroulement (figure 2 ).

Les systèmes I et II initialement séparés sont envoyés dans une zone
d'interaction où une fonction d'onde composée particulière {\Large
$\Psi_{I,II}$} est produite qui laisse les systèmes I et II se séparer
de nouveau. Insistons que les systèmes I et II se séparent mais la
fonction d'onde reste une tant qu'aucune mesure n'est effectuée sur
l'un ou l'autre des systèmes. Elle ne l'est plus ensuite, c'est ce
qu'on va voir. Le système I est dirigé, au choix, vers l'un de deux
appareils de mesure MSP pour mesurer $P_I$ ou MSQ pour mesurer $Q_I$,
P et Q étant des grandeurs conjuguées, ici position et impulsion. La
fonction d'onde {\Large $\Psi_{I,II}$} subit une ``réduction du paquet
d'onde'' (wave collapse dans EPR) dès que l'appareil MSP donne un
résultat $p_I$. Mais {\Large $\Psi_{I,II}$} est ainsi préparée qu'il
en résulte que II est lui aussi dans un état P ($p_{II}$ qui dépend de
la valeur $p_I$ trouvée sur I). De même, si c'est l'appareil MSQ qui
est présenté à I, des que l'appareil MSQ donne un résultat $q_I$, il
en résulte que II est dans un état Q ($q_{II}$ qui dépend de la valeur
$q_I$ trouvée sur I).

\includegraphics[angle=-90,scale=0.5]{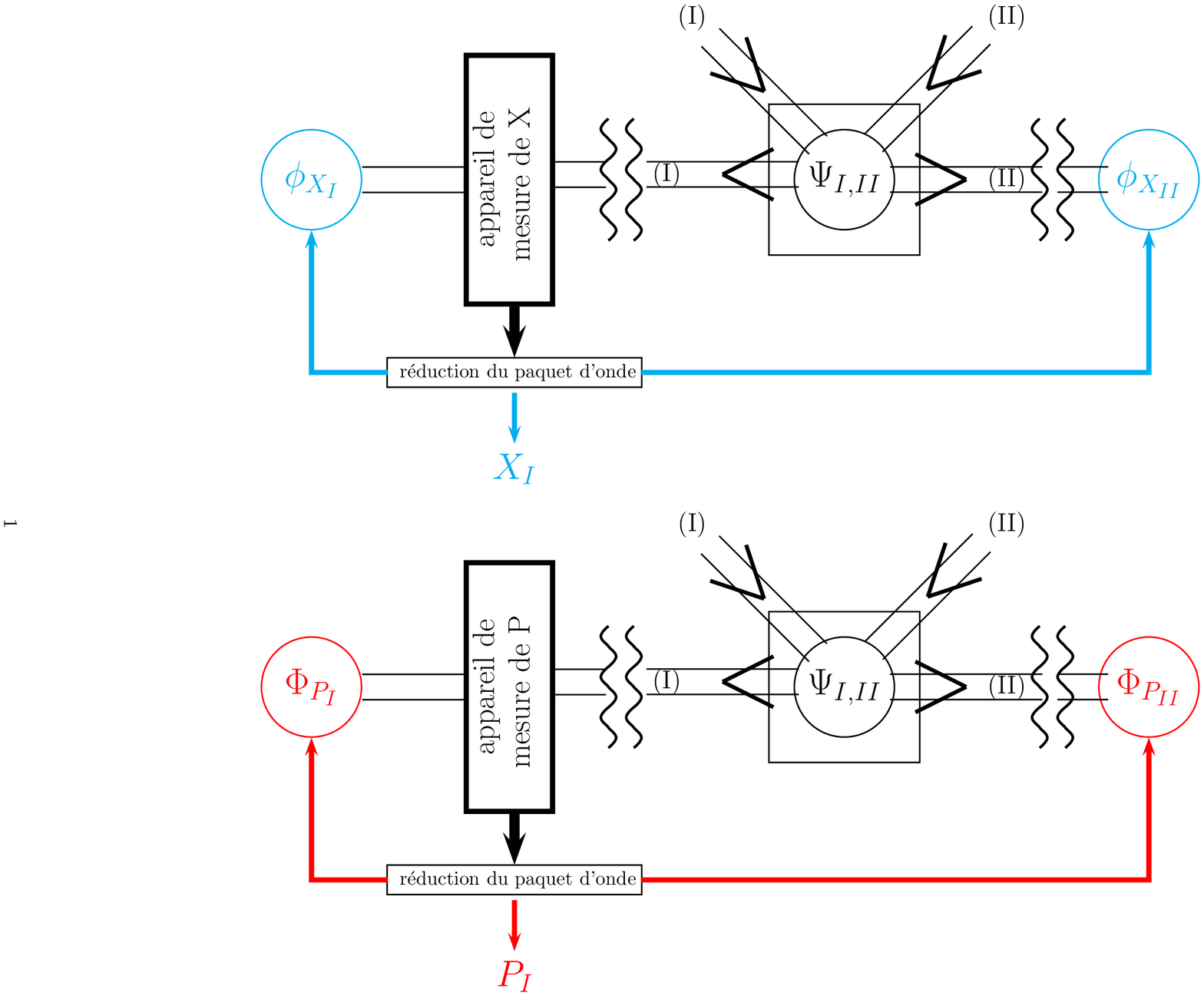}

EPR démontrent ainsi que l'état de II dépend d'une mesure effectuée
sur I avec lequel il n'interagit pas. Contradiction dans les termes
parfaitement démontrée dans le cadre de la MQ. Comment en sortir? Ou
bien II peut être à la fois dans un état P et un état Q, ce qui
contredit complètement la MQ : elle ne serait pas incomplète elle
serait incohérente. Ou bien la MQ introduit des ``actions à distance''
qui ne permettent plus d'affirmer, comme le supposent EPR :

\bq
{ \color{blue}  \sl\uline{ {\ldots}  we have two systems I and II , which we permit to
interact from the time t=0 to t=T, {\bf after which time \bf we suppose that
there is no longer any interaction between the two parts.}}}
\eq

C'est cette démonstration sous forme d'alternative qui est éternelle
{\ldots} à l'intérieur du cadre de la mécanique quantique, en 1935
comme aujourd'hui. Bien sûr que les choix dans l'alternative ou les
commentaires peuvent être différents hier et aujourd'hui!

\section{Tout simplement un dilemme.}

Pour EPR en 1935, clairement rien ne convient. Ni que p et q puissent
à la fois être déterminés, ni que des actions à distance soient alors
démontrées. Rejetant pourtant cette dernière possibilité, ils
affirment seulement :

\bq
{ \color{blue} \sl \uline{ We are thus forced to conclude that the quantum-mechanical
description of physical reality given by wave-functions is not
complete.}}
\eq

Ils se gardent bien par contre d'en déduire que P et Q puissent être
en même temps déterminés: ils ont trop confiance dans ce que prédit la
MQ. Voilà ce qu'en dira plus tard Einstein (en 1949, c'est vrai, bien
après 1935 oui, mais 15 ans avant 1964!) :
 
\bq
{ \color{blue}\sl\uline{ `` Cette théorie est jusqu'à maintenant la
seule qui unifie le double caractère corpusculaire et ondulatoire de
la matière d'une façon dont la logique est satisfaisante ; et les
relations (vérifiables) qu'elle contient, sont, à l'intérieur des
limites fixées par la relation d'incertitude {\sl complètes}. Les
relations formelles qui sont données dans cette théorie -c.a.d. son
formalisme mathématique tout entier- devront probablement être
contenues, sous la forme de conséquences logiques, dans toute théorie
future utile''{\normalfont\cite{eins1949} Einstein's reply, page
666-667. } }}
\eq

C'est peut-être pourquoi, voulant garder une distance avec le contenu
formel de sa démonstration, il n'en garde dans le titre qu'une mise en
cause générale, un questionnement même, sur la complétude de la
description quantique du monde physique.
 
Remarque : Si la MQ introduit des actions à distance, elle le fait
subrepticement, sans rien en dire, sans définir la nature de ces
interactions, leur portée finie ou infinie, leur vitesse de
propagation, finie ou infinie, leur articulation avec le cadre
existant de la relativité. Alors oui la MQ est incomplète.

\section{Ce que veut et ce que ne veut pas Einstein et la lecture
qu'en fait Bell}

On peut maintenant revenir à l'article de Bell, et on commence avec
l'introduction.

\bq
{\color{blue}\sl ``The paradox of Einstein, Podolsky, and Rosen was
advanced as an argument that quantum mechanics could not be a complete
theory but should be supplemented by additionnal variables. These
additionnal variables were to restore to te theory causality and
locality''}
\eq

On l'a vu, mécanique quantique incomplète oui, mais on chercherait en
vain dans le texte EPR le mot, l'idée même de variables
additionnelles, de paramètres cachés. Comment Bell peut-il écrire
cela?

Voici ce qu'en dira Einstein un peu plus tard :

\begin{quote}

{ \color{blue}\sl\uline{ Je ne pense pas que l'on puisse arriver à une
description des systèmes individuels simplement en complétant la
théorie quantique actuelle. Le principe de superposition et
l'interprétation statistique sont indissociablement liés entre eux. Si
l'on pense que l'interprétation statistique doit être dépassée, on ne
peut pas conserver l'équation de Schrödinger dont la linéarité
implique la superposition des états''}} Lettre à Kupperman de novembre
1953 (\cite{bali1989} page 233),

\end{quote}

C'était en 1953, 18 ans après EPR mais 11 ans avant Bell!

Revenons sur les termes de causalité et localité. Einstein est
clairement attaché à ce qu'on peut appeler la causalité (il n'emploie
pas lui même ce terme). On a souvent pappelé {\color{blue}\sl\uline{``Dieu ne joue pas au
dès''}}
et Einstein a toujours affirmé l'objectif de {\color{blue}\sl\uline {traiter ``les cas
individuels'}'} :

\bq
{\color{blue}\sl\uline{ ``Je suis en fait, et au contraire de presque
tous les physiciens théoriciens contemporains fermement convaincu que
le caractère essentiellement statistique de la théorie quantique
contemporaine doit uniquement être attribué au fait que cette théorie
opère avec une description incomplète des systèmes
physiques''}{\em\cite{eins1949}Einstein's reply, page 666. }}
\eq

Mais rien de cela n'apparait dans le texte EPR focalisé sur tout autre
chose.

Quant-au terme de localité lui aussi absent du texte EPR, il est
évoqué ici par Bell pour, en fait, introduire son contraire la
``non-localité''. Une façon par un mot de rendre compte de la
démonstration EPR mais sans en accepter la conclusion : la mécanique
quantique est incomplète. Quel est en effet le fondement scientifique
de cette ``non-localité''?

Un terme qu'on retrouve pourtant aujourd'hui encore et toujours aussi
peu scientifiquement fondé même s'il est en relation avec une
réalité aujourd'hui incontestable (tout aussi incontestable que la
vertu dormitive de l'opium, chère à Molière) de ou dans la mécanique
quantique. Une réalité mise à jour et démontrée en 1935 par EPR!

Bell poursuit :

\bq
{\color{blue}\sl ``In this note that idea will be formulated
mathematically and shown to be incompatible with the statistical
predictions of quantum mechanics''}
\eq

Comment peut-on formuler mathématiquement une idée qui n'est pas
présente (dans EPR)? Bell le tente dans la phrase qui suit :

\bq
{\color{blue}\sl ``It is the requirement of locality, or more
precisely that the result of a measurement on one system be unaffected
by opérations on a distant system with which it has interracted in the
past, that creates the essential difficulty''}
\eq

Mais exprimée comme cela, cette ``difficulté essentielle'' n'est-elle
pas celle de tout le monde?

On retrouve bien là le coeur de l'article EPR, mais de nouveau, ni
paradoxe ni difficulté, démonstration d'une action à distance,
non-localité si on veut, mais qu'aucun paramètre suplémentaire ne peut
ni faire disparaitre ni expliquer.

Ce que demontrent EPR, c'est que l'état de II (état P ou état Q) est
modifié par le déclenchement d'un appareil sur I et la réduction du
paquet d'onde qui y est associée, point n'est besoin d'une mesure sur
II qui, statistiquement donnera ce que prévoit la connaissance de
l'état II, quelle que soit cette mesure.

\section{La nécessité d'un nouveau participant : David Bohm}

Comment comprendre ces décalages (pour le moins) entre Bell et EPR
auquel il prétend se confronter? Cest qu'entre temps (1951-1957) en
effet, est apparue avec David Bohm une théorie quantique munie de
paramètres supplémentaires qui la rendent directement déterministe,
mais au prix d'actions à distance, de changements instantannés tout à
fait inconnus\cite{Brog52}. C'est ce que Bell écrit ensuite /

\bq
{ \color{blue}\sl ``Moreover, a hidden variable interpretation of
elementary quantum theory has been explicitly constructed. That
particular interpretation has indeed a grossly non-local structure.''}
\eq

Oui, cela peut bien reproduire les effets EPR, rappelons cependant que
les paramètres supplémentaires, locaux ou non ne sont pas ce qui
satisferait EPR en tout cas Einstein.

Bell termine son introduction :

\bq
{\color{blue}\sl``It is characteristic, according to the result to be
proved here, of any such theory which reproduces exacly the quantum
mechanical predictions''}
\eq

et Bell retourne  à Bohm, mais cette fois avec  Aharonov pour un
article\cite{bohm1957}
 directement en  relationa avec EPR  :

\bq
{\color{blue}\sl``With the example advocated By Bohm and Aharonov, the
EPR argument is the folowing''}
\eq

Mais quel est vraiment le rapport de Bohm et Ahronov avec EPR. Tout
EPR? Rien que EPR? La réponse n'est pas simple pas aussi simple que le
prétend Bohm lui même :

\bq
{\sl ``{\ldots} EPR have given an example of a hypothetical experiment
capable of testing certain apparently paradoxical predictions of the
current quantum theory. In order to illustrate this experiment we
shall consider a special example which permits us to present the
arguments of EPR in a simplified form. ''}
\eq

Car d'une part EPR ne proposent aucune expérience à vérifier et
d'autre part, comme on va le voir, on ne peut pas dire non plus que
l'expérience proposée par Bohm est une version simplifiée de
l'expérience de pensée EPR.

Mais reprenons plus en détail l'intervention de Bohm puisque c'est lui
qui est convoqué, et commençons par le titre :

\bq
{\sl Discussion of Experimental Proof for the Paradox of Einstein,
Rosen, and Podolsky }
\eq

Ce titre est inattendu puisqu'il focalise donc l'attention sur une
expérience rapportée dans la troisième partie de l'article et qui
démontrera que le paradoxe EPR (les actions à distance) est bien réel.
Mais cette expérience est menée avec des photons polarisés et pas avec
des particules de spin 1/2 comme proposé dans les parties 1 et 2 de
l'article. C'est à ces deux parties que se réfère Bell, pas du tout à
la partie expérimentale, celle qui justifie le titre ( celle aussi qui
d'une certaine façon peut sembler conclure le débat sur l'existence du
``paradoxe''!). Nous nous intéresserons nous aussi à ces deux parties.

\section{Bohm et EPR : quoi de commun, quelle diférence?}

Une molécule de spin total 0 est composée de deux atomes A et B de spin 1/2.

La fonction d'onde du système est alors :

\[ \psi = \frac{1}{\sqrt
2}[\psi_{+}(1)\psi_{-}(2) - \psi_{-}(1)\psi_{+}(2) ]     \]

Les deux atomes sont séparés par une opération qui conserve le spin.
Le spin d'un des atomes A est mesuré selon une direction quelconque et
la réponse + ou - de cette mesure permet de déduire celle trouvée sur
B si elle est conduite selon la même direction. L'état de B est
modifié par la mesure sur A. Il y a bien là équivallence avec EPR et
là aussi, le même raisonnement peut être déroulé : si une autre
direction est choisie pour la mesure sur A, alors B est aussi projeté
dans un état polarisé dans une autre direction alors que la MQ
interdit à un atome (B) d'être dans un état propre de spin sur deux
directions comme elle interdit avec EPR d'être en même temps dans un
état propre de P et de Q.

Bohm dit noter une difficulté spécifique à sa proposition (on peut
penser cependant qu'elle est transposable à EPR). La MQ attribue les
fluctuations de spin pour A sur les autres directions que celle
mesurée à l'interaction incontrôlable avec l'appareil de mesure sur A.
Mais il faudrait maintenant que cette interaction provoque aussi les
mêmes fluctuations sur B (``avec lequel il n'interagit pas''). On
retrouve les actions à distance.

Mais il y a pourtant une différence importante avec EPR : la
possibilité d'envisager une sortie du paradoxe par l'hypothèse d'un
mécanisme ad hoc, inventé pour la circonstance. Examinons ce que Bohm
propose et dont l'objectif est pour lui parfaitement clair :

\bq
{\sl {\ldots} There exists at present no experimental proof that the
paradoxical behavior described by ERP (sic) will really occur.}
\eq

Il attribue ensuite l'idée de la proposition qui suit à Einstein lui
même, dans une communication privée (Einstein est disparu depuis deux
ans) :

\bq{\sl namely,that the current formulation of the many-body problem in
quantum mechanics may break down when particles are far enough apart.}
\eq
Bohm reprend cette idée avec l'exemple qu'il utilise :

\bq
{\sl {\ldots} , we may consider {\ldots} that after the molecule of
spin zero decomposes, {\ldots} we suppose that in any individual case,
the spin of each atom becomes definite in {\normalfont some}
direction, while that of the other atom is opposite. The wave function
will be the product :}

\[ \psi = \psi_{+\theta,\phi}(1)\psi_{-\theta,\phi}(2)  \]

{\sl where $\psi_{+\theta,\phi}(1)$ is a wave function of particle A
whose spin is positive in the direction given by $\theta $ and $\phi
$}
\eq

Bohm indique alors que dans cette hypothèse, il n'y a plus
conservation du spin total pour un cas individuel, mais seulement en
moyenne!

\bq
{\sl {\ldots} ,but the model described above has the advantage of
avoiding the paradox of ERP}
\eq

Pas d'équivallent avec EPR, au moins pas d'équivallent qui ne remette
en cause l'ensemble de l'édifice de la MQ. Au mieux, il faudrait en
effet, pour EPR, que I et II se séparent emportant avec eux la double
information $x$ et $p_x$ ce qu'interdit radicalement la MQ. 

Insistons sur le rôle essentiel dans EPR de la réduction du paquet
d'onde. A l'oeuvre sur I au moment de la mesure sur I de $x$ ou de
$p_x$ (en conformité avec le principe de complémentarité de Bohr),
elle opère aussi sur II, c'est là qu'est la surprise : la découverte
d'une complémentarité à distance! Pourrait-on dire. I et II sont
séparés mais la fonctiond'onde reste une. Là est la nouveauté, là est
le problème. On voit bien alors comment Bohm, dans la parie 2 de son
article, tente d'apporter une solution à cette situation nouvelle :
Les systèmes se séparent et la fonction d'onde aussi, elle ne reste
pas une. Proposition intéressante mais complètement hors de la
mécanique quantique. Il s'agit d'une invention originale,
l'introduction d'une auto-mesure en quelque sorte. On remarquera
pourtant contre cette invention que la mesure en mécanique quantique
résulte de l'interraction d'un micro-objet avec un élément (appareil
de mesure ou non) macroscopique complètement absent lors de cette
séparation des deux atomes.
 
Alors, Bohm, version simplifiée de EPR?

Si on reprend les trois parties de son article;

1) L'expérience de pensée avec la molécule de spin total zéro. Une
véritable équivallence avec EPR et toute la MQ et rien que la MQ.

2) La brisure spontanée de symétrie et l'introduction correspondante
d'un paramètre supplémentaire comme tentative de contourner les
conclusions de EPR au prix de l'invention d'un mécanisme hors de la MQ.

3) Une expérience avec polarisation de photons et qui confirme
l'existence du ``paradoxe EPR'' mais que Bell ignorera complètement.

Avec 1) comme avec EPR, nulle possibilité d'introduire un paramètre
supplémentaire.

Avec 2), il s'introduit naturellement mais en sortant radicalement de
la MQ.

On va voir que Bell met à l'épreuve une autre introduction de
paramètres supplémentaires mais elle aussi complètement en dehors de
la MQ. Bell ne fais pas appel à un mécanisme explicite comme Bohm. Ses
conclusions sont plus générales c'est vrai, mais hors de référence à
la physique (et pas seulement quantique).
 
\section{Bell dans le détail}

Mais ces considérations établies à partir des textes se
reflètent-elles dans les calculs utilisés par Bell pour établir ses
inégalités et dans le modèle à la base de ce calcul (le modèle qui
introduit les paramètres supplémentaires)? Evidemment
oui, comme on va le voir.

Le modèle d'abord.

Bell le fait se référer\footnote{On notera que Bell n'ignore pas ce
que défendait Einstein bien après 1935, et là en 1949, dans un
document plusieurs fois utilisé ici, dans cet article.} à une
déclaration d'Einstein en 1949(\cite{eins1949} page 85) :

\bq{\color{blue}\sl\uline{But on one supposition, absolutely hold fast :
the real factual situation of the system $S_2$ is independent of what
is done with the system $S_1$, which is spatially separated from the
former }}
\eq

On retrouve bien là l'affirmation que parler de deux systèmes {\bf
séparés} a un sens. Pour Einstein mais pour chacun de nous non?

Mais c'est Bell lui même, et pas du tout Einstein qui fait découler de
cette déclaration la nécessité de l'introduction d'un paramètre
supplémentaire $\lambda $ dans la préparation de l'état, pour {\bf
prédéterminer} résultat des deux mesures qui vont suivre. Bell le fait
après avoir supposé raisonnable que l'orientation d'un polariseur
n'influence pas le résultat de la mesure sur l'autre.

L'introduction de paramètes supplémentaires n'est pas du tout conforme
aux souhaits de Einstein (voir plus haut), elle est par contre bien
conforme (avec la prédétermination qui en résulte) avec l'hypothèse et
le modèle avancés par Bohm 2). C'est cette hypothèse que Bell
généralise en quelque sorte en s'affranchissant de toute référence à
un mécanisme sous-jacent (pas de rupture spontanée, pas d'auto-mesure
pour faire naitre le paramètre). Le paramètre supplémentaire dans le
passé commun est posé au départ sans que son existence soit associée à
un mécanisme connu ou inventé comme pour Bohm2). Comme pour Bohm2, par
contre, il n'y a pas avec Bell de réduction du paquet d'onde : rien ne
se produit au moment de la mesure sur A. Là est la rupture avec EPR et
avec la mécanique quantique.

Ainsi, Bell écrit :

\bq
{\color{blue}\sl The vital assumption is that the result B for particle 2 does not
depend on the setting $ \overrightarrow {a} $, of the magnet for
particle 1, nor A on B.}
\eq

Avec EPR, c''est au contraire la présence de l'appareil de mesure de P
ou celui de Q et le collapse qui suit qui déterminent le résultat sur
B. L'un ne peut pas être une extension de l'autre.

Poursuivons avec Bell :

\bq
{\color{blue}\sl The result A of measuring $ 
\overrightarrow{\sigma_1}.\overrightarrow{a} $ is then determined by
$\overrightarrow{a}$ and $\lambda $, and the result B of measuring $ 
\overrightarrow{\sigma_2}.\overrightarrow{b} $ is then determined by 
$\overrightarrow{b}$ and $\lambda $}
\eq

La valeur moyenne du produit des deux composantes
$\overrightarrow{\sigma_1}.\overrightarrow{a}$ et
$\overrightarrow{\sigma_2}.\overrightarrow{b}$ est alors :

\begin{equation}
 P(\overrightarrow{a},\overrightarrow{b})=\int d\lambda \varrho(\lambda)
 A(\overrightarrow{a},\lambda)B(\overrightarrow{b},\lambda)
 \end{equation}

où $\varrho(\lambda)$ est la distribution de probabilité de $\lambda$.

C'est cette expression que Bell va démontrer être incompatible avec la
valeur attendue pour la mécanique quantique :

\[ <
\overrightarrow{\sigma_1}.\overrightarrow{a}.\overrightarrow{\sigma_2}.\overrightarrow{b}>
= - \overrightarrow{a}.\overrightarrow{b}\]

On sait que l'incompatibilité nécessite que trois directions de
mesures soient utilisées. On sait que l'expérience confirme cette
incompatibilité. Mais ce n'est pas ce sur quoi nous portons notre
attention ici.

Revenons au point de départ du modèle mis à l'épreuve par Bell et
rapporté ci-dessus :

\bq
{\color{blue}\sl The result A of measuring $ 
\overrightarrow{\sigma_1}.\overrightarrow{a} $ is then determined by
$\overrightarrow{a}$ and $\lambda $, and the result B of measuring $ 
\overrightarrow{\sigma_2}.\overrightarrow{b} $ is then determined by
$\overrightarrow{b}$ and $\lambda $}
\eq

On voit très clairement l'impossibilité que ce modèle puisse
reproduire l'expérience de pensée de EPR. Avec eux en effet, c'est la
mesure sur I qui produit un changement d'état sur II, la détermination
de la nature d'un état (sur I aussi bien sûr). Rien de tel avec Bell.

On montre dans l'appendice B que la même formule (1) de Bell est
obtenue, au signe près, si on prétend rechercher un paramètre caché
$\lambda $ qui permettrait de déterminer les résultats de deux mesures
de polarisation successives sur le même atome. On se convaincra que là
c'est absurde et sans rapport avec la mécanique quantique. Avec le
modèle de de Bohm-Bell, on reste simplement sans rapport avec la mécanique
quantique.

\section{En résumé}

\begin{tabular}{|c|c|c|c|c|}
\hline
 & & $P(\overrightarrow{a},\overrightarrow{b})$ = 
$< \overrightarrow{\sigma_1}.\overrightarrow{a}.\overrightarrow{\sigma_2}.\overrightarrow{b}>
 $ &$ P(\overrightarrow{a},\overrightarrow{a})$ & $
 P(\overrightarrow{a},\overrightarrow{a_{\bot}})$ \\
\hline
\hline

Avec collapse & MQ (Bohm1 et EPR) & $  - \overrightarrow{a}.\overrightarrow{b} =
cos(\theta)$ & -1
& 0 \\ 
 \hline
 \multirow{3}{*}{Sans collapse}& Bohm2 &  $  -1/3
 \overrightarrow{a}.\overrightarrow{b} $ & -1/3 & 0 \\
 & Bell particulier & $ -1 + 2 \theta/\Pi $ & -1 & 0 \\
& Bell général & $ P(\overrightarrow{a},\overrightarrow{b})=\int d\lambda
\varrho(\lambda)
 A(\overrightarrow{a},\lambda)B(\overrightarrow{b},\lambda) $ 
 & -1 & 0 \\
 \hline
\end{tabular}

\vspace{0.5 cm}

On a rassemblé ces résultats dans le tableau I. Seule la MQ avec EPR
appliqué au cas Bohm1 comprend comme il se doit le recours à la
réduction du paquet d'onde à l'origine même du paradoxe. On a fait
ensuite figurer le modèle Bohm2 qui correspond à la rupture spontanée
de symétrie au moment de la séparation des composants. Ensuite deux
modèles relevant de la problèmatique de Bell : une dépendance générale
d'un paramètre supplémentaire non attachée à un mécanisme physique
supposé. Dans le premier modèle, une distribution spécifique du
paramètre et une dépendance des valeurs trouvées $
A(\overrightarrow{a},\lambda) $ et $ B(\overrightarrow{a},\lambda) $
de ce paramètre sont choisis. Dans le second et dernier modèle,
l'expression la plus générale où les dépendances explicites sont
abandonnées.
 
On a fait figurer pour ces quatre choix, la valeur moyenne $
<\overrightarrow{\sigma_1}.\overrightarrow{a}.\overrightarrow{\sigma_2}.\overrightarrow{b}>
= P (\overrightarrow{a},\overrightarrow{b})$ et les valeurs
particulières $ P (\overrightarrow{a},\overrightarrow{a})$ et $ P
(\overrightarrow{a},\overrightarrow{a_{\bot}})$. Bell montre alors
avec les deux dernières colonnes que la MQ reste compatible si on se
limite à ces seuls choix (un seul pour Bohm2). Bell montre que par
contre, l'expression $-1+\frac{2}{\Pi}$ obtenue avec les dépendances
explicites est fort différente de la MQ des qu'on s'éloigne de ces
conditions particulières ( $ \overrightarrow{a}.\overrightarrow{b} = 1
$ ou $ \overrightarrow{a}.\overrightarrow{b} = 0 $). 

Avec l'abandon de cette dépendance spécifique, quand il atteint donc
la plus grande généralité, le modèle déterministe avec paramètre
supplémentaire dans le passé commun peut s'approcher bien plus de la
MQ et il faut une ruse, en quelque sorte pour s'en distinguer : le
choix de trois direction de mesure est nécessaire et aboutit à
l'obtention des fameuses inégalités que viole donc la MQ.

Ce qui étonne, ce n'est pas cette violation mais bien qu'il faille
tout cet arsenal tant, tout de même, le modèle mis à l'épreuve est
éloigné de cette MQ et rappelons une fois de plus éloigné des
exigences d'Einstein! Ces inégalités confirment la démonstration
d'EPR, elles ne la contredisent pas.

\section{On fait le point}

Près de 30 ans séparent EPR de Bell, Bohm se situant à mi-course. Mais
quelles ont été les réactions immédiates à EPR? On doit bien sûr citer
Bohr qui réplique immédiatement avec le même titre, la même
interrogation devrait-on dire. Pas la même réponse, non! Et puis Furry
et surtout Schrodinger\cite{schr35} qui répond avec une grande honnèteté et avec
beaucoup de détails. Et jusqu'à Bohm (et quelquefois jusqu'à
aujourd'hui!) le contenu est le même. Le ton\footnote{Ce ton, celui de
la certitude, de la fermeture même, on le retrouve chez la plupart des
acteurs, on en trouvera quelques citations dans \cite{rous2006}} était donné par Bohr
tellement cité :

\bq
{\sl \ldots a viewpoint termed `complementarity'' is explained from which
quantum mechanical description of physical phenomena would seem to
fulfill, within its scope, all rational demands of completeness}
(\cite{Bohr35} résumé, page 696).

{\sl Such an argumentation [celle d'EPR] however, would hardly seem suited
to affect the soundness of quantum-mechanical description, which is
based on a coherent mathematical formalism covering automatically any
procedure of measurement like that indicated } (\cite{Bohr35}, page
696).
\eq

Mais analysé avec lucidité par Bohm en 1957 :

\bq
{\sl It is clear that in Bohr's point of view, no paradox can arise in the
hypothetical experiment of ERP. For the system of two atoms plus the
apparatus which is used to observe their spins is, in any case,
basically unseparable and unanalysable, so that the question of how
the correlations come about simply has no meaning.}
\eq

Mais ce qu'ont apporté EPR c'est que l'appareil de mesure des spins
est en fait deux appareils aussi éloignés l'un de l'autre que
souhaité. L'argument de Bohr s'en trouve considérablement affaibli, au
moins plus exposé au doute.

Si Bohm dit accepter le point de vue de Bohr, il précise cependant :

\bq
{\sl {\ldots}; but we differ, in that we suppose that this combined system
is at least conceptually analysable into components which satisfy
appropriate laws.}
\eq

Bohm exprime alors l'alternative, ou bien des actions à distance ou
bien une physique plus profonde dont la MQ ne serait que
l'approximation.

Bohm termine cette partie 2 de son article, en avance sur ce qui va
venir en partie 3, et conclut avec la plus grande clarté\footnote{ En
1951\cite{bohm1951}, quand Bohm introduit le modèle des atomes couplés (la version
simplifiée de EPR!), il n'a pas encore ce point de vue et reste alors
un défenseur sans réserve du point de vue de Bohr} :

\bq
{\sl In sum, then, the quantum theory of the many-body problem implies the
possibility of a rather strange kind of correlation in the properties
of distant things. As we shall see in the next section, experiments
proving the presence of this kind of of correlation already exist.
Any attempt to interpret (sic) the quantum mechanics and to understand
its meaning must therefore take such correlations into account.}
\eq

Nous sommes en 1957. Il y a donc alors pour Bohm des corrélations
surprenantes dont la compréhension reste à construire.

Peut-être comprend-on alors pourquoi Bell en 1964 s'intéresse
principalement à la partie 1 de l'article de Bohm, celle qui lui
fournit le modèle à mettre à l'épreuve, un peu à la partie 2 pour
l'éliminer car trop éloignée de la MQ, et pas du tout à la partie 3.
Bohm conclut déjà dans cette partie 3 ce que Bell va vouloir encore
mettre à l'épreuve, encore le fait-il avec un modèle très général mais
sans grand rapport avec la MQ (oui, sans collapse).

\section {  Einstein et la mécanique quantique : quatre points forts.}

Au delà du rapport avec Bell, il peut être intéressant de schématiser
le point de vue d'Einstein sur la mécanique quantique dont il fut
,rappelons le, l'un des découvreurs majeurs

1) La description que fournit la mécanique quantique et les
prédictions qui s'en déduisent sont justes.
        
2) La mécanique quantique est insuffisante, inachevée, incomplète car
elle ne traite pas les cas individuels.
                
3) Pas de bricolage possible, pas de paramètres supplémentaires. Pour
arriver à la théorie complète il faut une refondation, une
reconstruction.
                            
4) Démonstration de l'alternative EPR et la question des actions à
distance.
                                    
On peut dire qu'on trouve la trace des trois premiers points tout au
long du demi-siècle qui sépare 1905 de 1955 ; pour le quatrième, c'est
plus compliqué. S'il est forcément schématique de rigidifier ainsi en
quatre points ce que furent cinquante ans d'interventions sur le
sujet, au moins, ce schéma est-il construit à partir de textes qu'on
peut vérifier! On en trouvera de plus larges extraits dans \cite{rous2006}.

Pour les points 2 et 3, on peut tenter un parallèle avec la
relativité. La mécanique classique pourrait être dite incomplète, mais
on ne passe pas à la mécaniqque relativiste en ajoutant des paramètres
supplémentaires. La mécanique classique devient une approximation de
la mécanique relativiste. Des situations expérimentales distinguent
l'une de l'autre. Mais pour la MQ, il suffirait que la mécanique du
futur englobe la MQ en éliminant, même seulement en principe, le
caractère probabiliste. Voilà le genre de considérations que proposait
Einstein sur le sujet :

\bq

{\color{blue}\uline{\sl``Il me semble en tout cas, que l'alternative
continu-discontinu est une authentique alternative ; cela veut dire
qu'ici, il n'y a pas de compromis possible. {\ldots} Dans cette
théorie, il n'y a pas de place pour l'espace et le temps, mais
uniquement pour des nombres, des constructions numériques et des
règles pour les former sur la base de règles algébriques excluant le
processus limite. Quant à savoir quelle voie s'avérera la bonne, seule
la qualité du résultat nous l'apprendra''}} Lettre à Joachim,
\cite{bali1989} page 256.
\eq 
                                 
Concernant les actions à distance, voilà ce qu'il en dit en 1949 :

\bq
{\color{blue} \uline{\sl I close these expositions, which have grown
rather lengthy, concerning the interpretation of quantum theory with
the reproduction of a brief conversation which I had with an important
theoretical physicist.}}

{\color{blue}\uline{\sl 
He : ``I am inclined to believe in
telepathy.''}}

{\color{blue}\uline{\sl
I : `` This has probably more to do with physics than with psychology.''
}}

{\color{blue}\uline{\sl
He : ``Yes''
}}
\eq
 \cite{eins1949} page 683.

\section{Plus de questions que de réponses, plus de doute aussi.}

Einstein est un des artisans majeurs de la consruction de la MQ.
Pourtant, en 1935 il y est marginalisé. Pourquoi alors est-ce lui qui
fait la découverte de ces corrélations, ce phénomène physique nouveau,
et pourquoi sa découverte est-elle ignorée{\ldots} jusqu'à Bohm en
1957 et Bell en 1964? Mais aussi, pourquoi pas les autres?

\begin{wrapfigure}[11]{r}[34pt]{10 cm}
\includegraphics[ scale = 0.9]{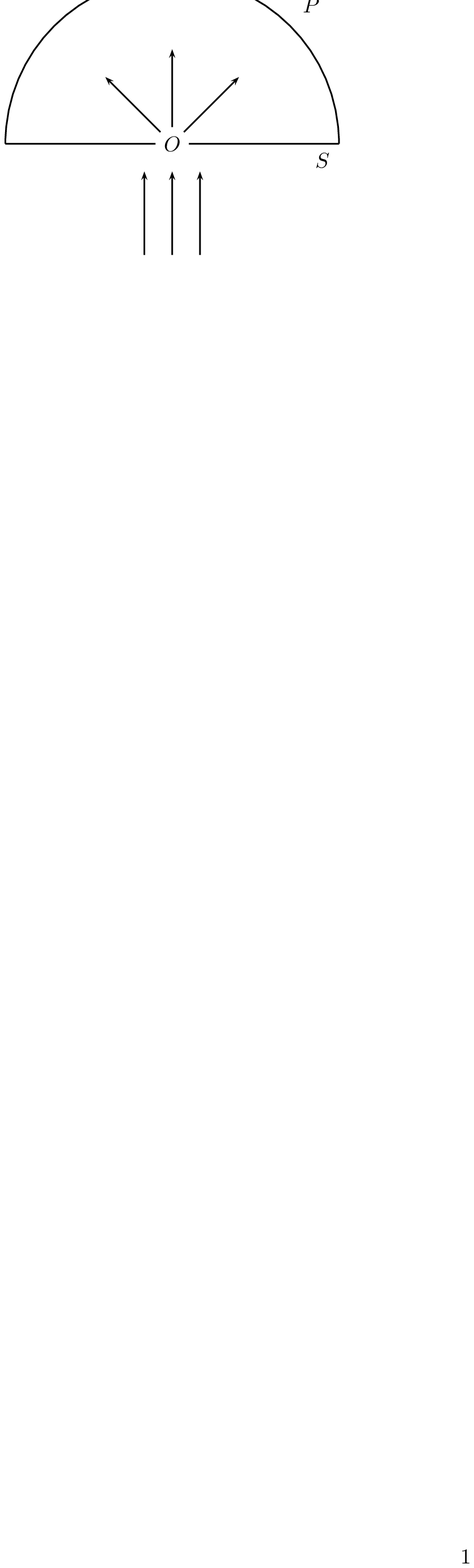}

\end{wrapfigure}

Einstein, depuis le début, est préoccupé par le statut de la fonction
d'onde. Il est toujours tenté d'y accorder une valeur statistique
sinon on a un problème avec la relativité au moment du collapse. Il a
souvent utilisé l'expérience de pensée (une de plus!) très simple,
schématisée sur le dessin ci-contre (figure 3). Une particule ou un
photon sont diffractés au passage d'un petit orifice $O$ dans un écran
$S$ puis détectés sur une plaque photographique semi-sphérique $P$. Et
son commentaire (ici au cogrès Solvay de 1927) :

{ \color{blue}\sl\uline{Mais l'interprétation d'après laquelle $|\Psi|^2 $ exprime la
probabilité que {\em cette} particule se trouve à un endroit déterminé
suppose un mécanisme d'interraction à distance tout particulier, qui
empèche que l'onde continuement répartie dans l'espace produise une
action en {\em deux} endroits de l'écran.}}

Quelle est la réponse à cette question complètement légitime dés
qu'une part de réalité est donnée à la fonction d'onde? C'est
justement de dire que la fonction d'onde n'est pas réelle (un état
dans l'espace de Hilbert). Schrodinger dont l'expression est la plus
détaillée et la plus honnète dit que c'est ``le catalogue des
réponses possibles de la mesure avec leur probabilités''. La réponse
c'est aussi de dire avec Bohr (comme le rappelle bien Bohm) qu'un
ensemble matériel n'est pas analysable en ses parties {\ldots} tant
que la mesure n'est pas accomplie. On ne peut pas, on ne veux pas
savoir ce qui se passe mais seulement quel sera le résultat. On fuit
la physique pour se réfugier dans l'interprétation\footnote{On ne peut
pas non plus passer sous silence les décennies pendant lesquelles,
plus ou moins clairement affirmé, un rôle a été attribué à
l'observateur, quand ce n'était pas à sa conscience (voir de
nombreuses citations dans \cite{rous2003}).}!

On voit bien comment EPR rendent cette position un peu plus difficile,
c'est exprimé clairement par Bohm, 22 ans après, on l'a vu plus haut.
Et la réponse ne sera pas l'abandon de cette posture (on ne veut rien
savoir) mais le refuge dans un mot la non-localité, expression
complètement absente avec Bohr et aujourd'hui complètement intégrée
depuis -au moins- Bell. On doit remarquer, aussi insuffisant soit-il,
qu'avec ce mot on se rappoche tout de même de la réalité, de la
physique, on veut déjà voir, avant de savoir.

\section{En guise de conclusion}

En 1935, Einstein et ses collègues Podolsky et Rosen démontrent
l'existence d'une propriété insolite/étonnante/inattendue de la MQ et
s'interrogent alors sur sa cohérence globale, sa complétude.  De quoi
s'git-il. Lorsque les deux éléments d'un sytème composite
convenablement préparé se séparent, la fonction d'onde reste une et
peut conduire à la production de corrélations au moment où des mesures
et le collapse correspondant sont réalisées. Cette découverte est
restée inaperçue, noyée dans les concepts flous de complémentarité et
de non-séparablité. Il faudra attendre 22 ans (le monde il est vrai
sera bouleversé dans cet intervalle) et la disparition d'Einstein pour
que, avec Bohm, l'attention soit portée sur cette découverte comme
découverte d'un phénomène physique exigeant une conpréhension
spécifique.

Si sa compréhension n'a pas alors avancé, au moins un mot a-t-il été
inventé pour le décrire : non-localité, un peu plus explicite et
spécifique d'un phénomène physique [que l'ancienne non-séparibilité].

Avec Bohm [selon peut-être une idée d'Einstein], est examinée la
possibilité d'éviter le paradoxe/phénomène en modifiant/complètant la 
MQ d'une manière radicale par l'intervenntion d'une brisure spontanée
de symétrie au moment (lequel?) de la séparation des composants. Cette
brisure aboutissant à un paramètre supplémentaire dans le passé
presque-commun, la directon commune de polarisation. Les résultats de
cette hypothèse sont très différents de ceux de la MQ comme le
montrera plus tard Bell. 

Mais Bohm affirme clairement alors et en s'appuyant sur des résultats
expérimentaux, que le phénomène d'action à distance, le paradoxe est
bien réel.

En 1964, Bell reprend pourtant l'idée de paramètres cachés dans le
passé commun mais en en généralisant la base, plus de mécanisme
générateur de ce paramètre, faiblesse et force de son hpothèse. Il
montre alors qu'un tel modèle dans ses résultats peut se rapprocher
beaucoup plus de la MQ que le modèle de Bohm puisque des inégalités
sont nécessaires impliquant trois directions de mesure pour les
différentier.  Il reste que ce modèle, malgré la proximité de ses
résultats avec ceux de la MQ est tellement éloigné de celle-ci qu'on
ne peut en aucun cas le considérer comme une extension de celle-ci. On
ne peut non plus le considerer comme une réponse à l'article original
EPR tant dans celui-ci la mesure/le collapse est essentiel au paradoxe
démontré et complètement absent du modèle de Bell.

C'est pourtant cet article de Bell et les inégalités qu'il établit qui
vont alimenter expériences et commentaires sur ce sujet pendant des
décennies et jusqu'à aujourd'hui.

Et pour revenir à la question évoquée au début de cet article, il
semble bien difficile de faire entrer l'enchainement des arguments
échangés durant cette longue période dans l'histoire qu'on raconte!

\section*{Appendice A : Mesure, collapse ou réduction du paquet d'onde?}

Rien n'est vraiment satisfaisant : mesure peut laisser croire à un
rôle pour l'observateur, collapse ou réduction du paquet d'onde peut
laisser croire que tout est dans la représentation, dans l'espace de
Hilbert. Le vocabulaire est ambigu parce que la chose elle même n'est
pas aussi claire qu'on le souhaiterait. Et puis cela résonne avec les
questions délicates de la dualité onde/corpuscule ou celles concernant
le statut de la fonction d'onde.

Gardons collapse, plus spécifique que réduction, plus court si on veut
joindre ``du paquet d'onde''et qui peut marquer le caractère concret
et objectif d'un processus. On peut garder mesure lorsqu'il y a
détermination de la valeur d'un paramètre avec ou sans présence d'un
observateur.  

\newpage
\vspace{1.5cm}
\large
\hspace{4cm}\fbox{\parbox{5cm}{
\centerline{\bf Mesure donnant}

\centerline{\bf le résultat {\sl a$_n$}}}}

\begin{pspicture}(16,5)

\psline[linewidth=.06]{c->}(6.6,5)(6.6,4)
\psline[linewidth=.06]{c-}(6.5,4)(6.5,0)
\psline[linewidth=.06]{c-}(6.7,4)(6.7,0)
\psline[linewidth=.05]{c->}(0,0)(13.5,0)
\psline[linewidth=.05]{c-}(1,.3)(1,-.3)
\psline[linewidth=.05]{c-}(11.75,.3)(11.75,-.3)
\rput{0}(13.7,0){{\em t}}
\rput{0}(11.75,-.5){{\em t$_1$}}
\rput{0}(6.6,-.5){{\em t$_0$}}
\rput{0}(1,-0.5){0}
\rput[bl]{0}(0.,1){$\ket{\psi(0)}$}
\rput[bl]{0}(4.8,1){$\ket{\psi(t_o)}$}
\rput[bl]{0}(6.9,3){$\ket{u_n}$}
\rput[bl]{0}(11.,3){$\ket{\psi'(t_1)}$}

\multips{0}(1.8,1)(.8,0){3}{
\pscurve[linewidth=.05](0,.15)(0.2,0)(.4,.15)(.6,.3)(.8,0.15)
}
\psline[linewidth=.05]{c->}(4.2,1.15)(4.5,1.15)

\multips{0}(8.,3)(.8,0){3}{
\pscurve[linewidth=.05](0,.15)(0.2,0)(.4,.15)(.6,.3)(.8,0.15)
}
\psline[linewidth=.05]{c->}(10.4,3.15)(10.7,3.15)
\end{pspicture}
\vspace{7.mm}
\large

\noindent FIGURE 2

{\bf Lors d'une mesure à l'instant {\sl t$_0$} de l'observable {\sl A}
donnant le résultat {\sl a$_n$}, le vecteur d'état du système subit une
brusque modification, et devient $\ket{u_n}$. Il évolue ensuite à partir de
ce nouvel état initial. }
\vspace{9mm}

La figure 2, reproduite de la référence \cite{Cohe77} page 221,
présente la question de la {\sl mesure} . Le vecteur d'état (la
fonction d'onde) évolue de façon déterministe (équation de
Schrödinger) depuis la préparation initiale en {\em t} = 0 jusqu'à la
{\bf mesure} en {\sl t = t$_o$}. Il subit alors un changement brusque
probabiliste vers l'un des vecteurs propres $\ket{u_n}$ de A, celui
qui est associé à la valeur propre a$_n$ trouvée. Le vecteur d'état
(la fonction d'onde) est projeté sur un de ses vecteurs propres et
renormalisé. Il reprend ensuite une évolution déterministe depuis {\sl
t = t$_o$} jusqu'à (par exemple) {\sl t = t$_1$} où une nouvelle
mesure est éventuellement pratiquée etc\ldots

Cette présentation n'est pourtant pas suffisante et pour deux raisons :

\begin{itemize}
\item 
ne sont pas du tout évoquées (ce n'est pas si simple, c'est vrai!) les
circonstances qui font qu'on passe d'un comportement à l'autre. On
précisera simplement nous, qu'un objet macroscopique éventuellemnt
partie d'un appareil de mesure, vient interagir avec la particule
microscopique à laquelle est associée la fonction d'onde.  Insistons
pourtant que la plupart des réductions ont lieu dans l'univers sans
qu'aucun observateur ne soit présent et qu'il est donc complétement
exceptionnel que cela se produise dans un appareil de mesure.

\item
n'est pas noté non plus que souvent (toujours?) la mesure d'un
paramètre correspond à la localisation de la particule dans un
détecteur ou une partie d'un détecteur et la fonction d'onde d'espace
est également modifiée au moment de la mesure recherchée et décrite
dans la figure 2 d'un paramètre autre que la localisation.

\end{itemize}

On a parlé de localisation de la particule, il faudrait mieux dire
réduction de la localisation car on n'arrive jamais à une localisation
ponctuelle, elle n'a pas de sens. C'est cette réduction qui se produit
si souvent dans l'univers en dehors de tout apparail de mesure. Alors
oui, pourquoi ne pas parler de {\sl \guillemotleft collapse
\guillemotright} pour le phénomène général et garder {\sl
\guillemotleft mesure \guillemotright} ... quand il y a mesure c'est à
dire détermination de la valeur d'un paramètre par ce {\sl
\guillemotleft collapse \guillemotright}. 
 
Mais justement, dans l'examen qui précède des articles EPR et Bell on
a beaucoup parlé de mesure, cette réduction/localisation, ce collapse
très exceptionnel.

\section {Appendice B : spin, MQ et paramètres supplémentaires.}

On examine dans ce qui suit avec un petit peu plus de détails les
propriétés du spin dans la MQ et des possibilités de rendre compte des
mêmes résultats avec un modèle (classique ) à paramètres
supplémentaires.

\subsection*{Une simple mesure de spin}

On considère une particule de spin 1/2. 

Elle est préparée dans un état pur de polarisation selon la direction
d'un vecteur unitaire $\overrightarrow{p}.$ La mesure
$\overrightarrow{\sigma}.\overrightarrow{a}$ de la polarisation de
cette particule selon une direction $\overrightarrow{a} $ va donner +1
ou -1 avec une valeur moyenne :

\[  < \overrightarrow{\sigma}.\overrightarrow{a} > = cos(\theta) \]

où $ \theta$ est l'angle entre$ \overrightarrow{p}$ et
$\overrightarrow{a}.$

La MQ en effet ne permet pas de prédire la valeur trouvée -1 ou +1
mais seulement sa valeur moyenne.

Si $\overrightarrow{a}$ est identique à $\overrightarrow{p}$, l'état
de polarisation est inchangé. Avec $\overrightarrow{a}$ différent de
$\overrightarrow{p}$, si la mesure a donné +1 l'état est polarisé
selon $+\overrightarrow{a}$, si elle a donné -1 l'état est polarisé
selon $-\overrightarrow{a}$. Cela est bien conforme à la bonne mesure!

On est prêt à faire une seconde mesure.

\subsection*{une mesure double ou deux mesures?}

Supposons donc qu'après la première mesure une nouvelle mesure est
effectuée selon une direction $\overrightarrow{b}$, différente de
$\overrightarrow{a}$, et que la première ait donné par exemple +1. De
nouveau, le résultat selon $\overrightarrow{b}$ sera +1 ou -1 de façon
inprédictible sauf en moyenne puisque cette fois,

\[  < \overrightarrow{\sigma}.\overrightarrow{b} > = cos(\theta') \]

où $ \theta'$ est l'angle entre$ \overrightarrow{a}$ et
$\overrightarrow{b}.$ (si le résultat de la mesure sur
$\overrightarrow{a}$ avait été -1 au lieu de 1, $ \theta'$ aurait été
l'angle entre$ -\overrightarrow{a}$ et $\overrightarrow{b}.$)

Notons que, pour les moyennes au moins, le résultat dépend de $
\overrightarrow{a}$ et de $\overrightarrow{b}$ mais plus du tout de
$\overrightarrow{p}$ dont on peut dire que le souvenir a été en
quelque sorte effacé par la première mesure.

Pour la MQ, le processus de mesure avec sa projection/renormalisation
est tel que chaque mesure donne au système un recommencement.

\subsection*{On complète la MQ? Introduction de paramètres supplémentaires}

Peut-on imaginer, au delà de la MQ donc, qu'un paramètre $\lambda$
dont on ignorerais la valeur déterminerait le résultat de la
réduction/localisation (simple localisation ou la valeur trouvée pour
une mesure) pour chaque événement.
La valeur prédite (avec certitude!) pour la moyenne par la MQ étant respectée
grace à la distribution particulière
 $D(\overrightarrow{\lambda})$ des valeurs de $\lambda$, chaque résutat possible
de la mesure étant associé à une partie de
la distribution $D(\lambda)$

Cette distribution $D(\overrightarrow{\lambda})$ doit évidemment
dépendre de l'état intial et d'autre part la règle qui permet de
déterminer le résultat doit dépendre de la disposition particulière de
l'objet macroscopique qui a provoqué la réduction/localisation, et
pour une mesure, du choix de l'appareil et de la valeur des paramètres
qui peuvent le caractèriser. Mais ce qui semble absolument nécessaire
pour être conforme à la MQ et à son processus de mesure
(projection/renormalisation), c'est que la distribution D($\lambda$)
est réinitialisée par chaque mesure.

Précisons dans le cas d'une mesure de spin telle qu'elle est évoquée plus haut.

La distribution $D$ dépend de $\overrightarrow{p}$, le vecteur selon
lequel la dernière mesure a été effectuée et de $\overrightarrow{a}$
celui qui va maintenant servir, la présence de l'appareil de mesure en
quelque sorte.

Alors, peut-on compléter la MQ et trouver cette distribution
$D(\overrightarrow{\lambda})$ qui réponde à nos objectifs? Pour une
mesure simple (sur un état initialement polarisé) la réponse est oui.
Oui c'est possible, cela ne dit pas bien sûr que c'est la réalité.
C'est possible. Montrons que c'est possible, qu'on peut définir une
distribution $D(\lambda)$ qu'on va pouvoir partager entre
$D^+(\lambda)$ pour lequel le résultat de la mesure est +1 et
$D^-(\lambda)$ pour lequel il est -1.

\subsection*{On commence par la mesure simple}

On définit d'abord la distribution 
$D(\overrightarrow{\lambda};\overrightarrow{p})$ une distribution
uniforme de $\overrightarrow{\lambda}$ sur la demi sphère 
$\overrightarrow{\lambda}.\overrightarrow{p} \leq 0$

On va maintenant séparer cette distribution en ses deux sous-ensembles
$D^+(\lambda)$ et $D^-(\lambda)$. Soit $\overrightarrow{a'} $ un
vecteur unitaire dépendant de $\overrightarrow{a}$ et de
$\overrightarrow{p}$ de telle sorte que :

\begin{enumerate}
 \item
le résultat de la mesure $ \overrightarrow{\sigma}.\overrightarrow{a}=
signe(\overrightarrow{\lambda}.\overrightarrow{a'} )$ est déterminé
par la valeur de $\overrightarrow{\lambda}$
\item
la moyenne de ce résultat est conforme aux prédictions de la MQ .
\end{enumerate}

Si $\Theta '$ est l'angle entre $ \overrightarrow{a'}$ et
$\overrightarrow{p}.$ Si $\Theta $ est l'angle entre $
\overrightarrow{a}$ et $\overrightarrow{p}.$

On doit avoir :

\[ 1- \frac{2\Theta'}{\Pi} = cos(\Theta) \]

C'est l'angle $\Theta '$ entre $ \overrightarrow{a'}$ et
$\overrightarrow{p}$ qui est ainsi déterminé (une infinité de
directions répondent à cette condition.)

C'est John Bell qui a proposé ce petit modèle. Von Neumann avait
pourtant démontré que des paramètres cachés étaient incompatibles avec
la MQ mais c'est aussi John Bell qui a montré que la supposée
démonstration de Von Neumann nécessitait une hypothèse que Bell a
montré être sans fondement.

\subsection*{... et la double mesure?} La remarque faite plus haut
s'applique directement : la première mesure a fait ``oublier'' l'état
initial, c'est le processus essentiel de projection/normalisation et
les nouveaux paramètres ne peuvent être communs avant ou après cette
première mesure. Inutile donc, absurde même de chercher un modèle qui 
permettrait à partir d'une distribution $D(\overrightarrow{\lambda})$
de prévoir le résultat de la mesure sur $\overrightarrow{p} $ puis sur
$\overrightarrow{a}$... et pourquoi pas sur toute autre direction, la
suite infinie des mesures possibles, ... futures et passées! . On
voudrait de plus que la valeur moyenne $<
\overrightarrow{\sigma}.\overrightarrow{p}.\overrightarrow{\sigma}
.\overrightarrow{a} > = \overrightarrow{a}.\overrightarrow{p}$ que
prévoit la MQ soit respectée.

Inutile, absurde, mais est-ce possible?

Formalisons la question : Le résultat A de la mesure
$\overrightarrow{\sigma}.\overrightarrow{a}$ est déterminé par$
\overrightarrow{a}$ et $\lambda$.

Le résultat B de la mesure
$\overrightarrow{\sigma}.\overrightarrow{b}$ est déterminé par
$\overrightarrow{b}$ et $\lambda$.

\begin{equation}
 A(\overrightarrow{a},\lambda)=\pm 1, B(\overrightarrow{b},\lambda)= \pm 1
\end{equation} 

et la moyenne des résultats obtenus est conforme aux prédictions de la MQ.

\begin{equation}
 P(\overrightarrow{a},\overrightarrow{b})=\int d\lambda \varrho(\lambda)
A(\overrightarrow{a},\lambda)B(\overrightarrow{b},\lambda)
\end{equation} 

On constate alors que les équations 1 et 2 ci-dessus sont les mêmes
(au signe près pour la seconde) que celles utilisées par John Bell
dans l'article examiné plus haut. Est-ce une surprise? Non, dans les
deux cas, pas de réduction du paquet d'onde!

\end{document}